\documentclass{aa}
\usepackage{graphics}

\begin{document}
\input psfig.sty

\thesaurus{03(04.19.1; 11.01.2; 13.18.1)}
\title{The $408$ MHz B3.1 Survey}
\author{M. Pedani \inst{1,2}, G. Grueff \inst{1,2}}
\institute{Dip. di Astronomia, Universit\'a di Bologna, Via Ranzani 1,
I-40127 Bologna, Italy
\and Istituto di Radioastronomia del CNR, Via Gobetti 101, I-40129
, Bologna, Italy}

\titlerunning{The 408 MHz B3.1 Survey}
\authorrunning{Pedani, M., Grueff, G.}
\offprints{M. Pedani; pedani@ira.bo.cnr.it}
\date{Received 25 May 1999/ Accepted 26 July 1999}

\maketitle
\begin{abstract}
We present a new $408$ MHz survey (B3.1) carried out with the 
''Croce del Nord'' radiotelescope in Bologna. 
The survey  coordinates limits are $-2^\circ 00^\prime$ to $+2^\circ
15^\prime$ 
in Dec. and $21$h to $24$h, $00$h to $17$h in R.A., equivalent to 
$0.388$ sr. 
The B3.1 is complete to $0.15$ Jy but many sources down to $0.1$ Jy
are included. Our aim was to select 
a new and complete sample of Ultra Steep Spectrum (USS) radio sources, as 
they proved to be good candidates to find high-$\it z$ radiogalaxies 
and their surrounding protoclusters. 
The observations and the reduction procedure are described and the 
observational errors are discussed. 
A cross-identification with the NVSS survey was performed to obtain the 
spectral index 
$\alpha_{408}^{1400}$ and radio size of the sources. We found no evidence of a 
change of the spectral index distribution as radio flux decreases. 
The B3.1 USS sample contains $185$ sources down to $0.1$ Jy and 
it is about one order of magnitude deeper in flux with respect to the 
4C USS sample. 
For $146$ B3.1 USS sources no optical counterpart was found on the
POSS-I sky survey. 
A cross-correlation with the 
FIRST survey gave maps for a subset of $50$ USS sources, and their optical
ID search was also made on the POSS-II, resulting in $39$ empty fields. 
\end{abstract}
\section{Introduction}

The study of powerful high-z radiogalaxies ( hereafter $HzRGs$ ) 
can yield crucial information on the properties, on the formation epoch 
and on the evolution of radiosources and their surrounding protoclusters.
To this end it is essential to have statistically homogeneous samples.
The Ultra Steep Spectrum sources ($S_{\nu} \propto \nu^{\alpha}$,
somewhat depending on the selection frequency, $\alpha \leq -1$; hereafter
 USS) of relatively small angular size have been proved over the past 
years to be good tracers of powerful $HzRGs$. 
Up to the last year there were about $120$ radiogalaxies known with 
$z > 2$ (de Breuck et al. 1997)
, but these have been obtained from collection of different 
datasets, resulting in quite heterogeneous samples, from which it is 
difficult to derive statistical answers (as stressed by R\"{o}ttgering 
et al. 1997). To date, the most distant known AGN is a 
radiogalaxy with redshift $z = 5.19$ ( van Breugel et al. 1999 ). 
The radio source was selected from a new sample of ultra-steep spectrum 
sources and it has an extreme radio spectral index of $\alpha_{365}
^{1400} = -1.63$. 
For a detailed description of the USS properties 
and of some existing samples see R\"{o}ttgering et al. (1994). 
With the ''Croce del Nord'' radio telescope in Bologna (see Ficarra 
et al. 1985 for the instrument description) we have performed a new 
$408$ MHz survey, down to $0.1$ Jy, tailored to provide an 
homogeneous and low-frequency selected sample of USS. The survey 
was made at the lowest declination observable with the ''Croce del Nord''. 
The sky area has been chosen so to be easily observed both by
northern and southern emisphere optical telescopes. An identical 
search (Rhee et al. 1996) was carried out over this sky region to select 
4C USS but, if we assume a spectral index $\alpha = -1$, the B3.1 USS 
sample is about one order of magnitude 
deeper in flux with respect to the 4C USS. Presently it is not known whether 
this lower flux limit will result in an increased fraction of 
$HzRGs$ in our survey with respect to the 4C.

\noindent
The NVSS (Condon et al. 1998) maps have been used to 
find the $1.4$ GHz counterpart of each $408$ MHz selected source and to 
compute the spectral index $\alpha_{408}^{1400}$. 

\noindent
The present paper provides a description of this Bologna sky survey 
( to be referred as the B3.1 survey ) and of the new USS sample 
(B3.1 USS). 
The survey and the data are described in Sect. 2. 
Section 3 deals with the VLA-NVSS maps cross-identification. 
Consistency and data quality checks are discussed in Sect. 4. 
In Sect. 5 the B3.1 USS sample radio properties are described, together 
with preliminary results of the optical identification programme.

\smallskip
\section{The B3.1 Survey: observations and data}
The surveys performed with the ''Croce del Nord'' radio telescope 
prior to the B3.1 are summarized in Table 1. 

\begin{table*}
\begin{center}
\caption[]{The radio surveys performed with the ''Croce del Nord'' 
radiotelescope at $408$ MHz. In the last column, the abbreviations are as 
follows: Bra65 = Braccesi et al. (1965); GV68 = Grueff \& Vigotti 
(1968); 
Co70 = Colla et al. (1970); Co72 = Colla et al. (1972); Co73 = Colla et
al. (1973); Fa74 = Fanti et al. (1974); Fi85 = Ficarra et al. (1985)}
\begin{tabular}{ccccccl}
\hline
Survey & Lim. Flux (Jy)& \# sources& $HPBW$ (arcmin) & Sky coverage & 
Year & Ref. \\
\hline
       &           &           &               &      &   &    \\
B1    & $1$ & $654$ & $4 \times 108$ & $3h 00m < R.A. < 13h 00m (1950)$ &
 $1965$ & Bra65  \\
   &  &  & & $-30^{\circ}< \delta < -20^{\circ} (1950)$ &   &   \\
       &               &               &         &       &      &   \\
GV     & $0.15$& $328$ & $3 \times 10$ & $7h 40m < R.A. < 18h 20m (1950)$ 
& $1968$ & GV68 \\
       &       &        &  & $+34^{\circ}< \delta < +35^{\circ} (1950)$  
 &  &  \\
       &       &    &    &   &    &  \\
B2     & $0.2$        & $3235$  & $3 \times 10$ & $+29^{\circ}
18^{\prime} < \delta < +34^{\circ} 02^{\prime} (1968)$ & $1970$ & Co70 \\
       &              &             &              &         &    &  \\
B2.2   &  $0.25$      & $3013$     & $3 \times 10$ & $+24^{\circ} 
02^{\prime} < \delta < +29^{\circ}30^{\prime} (1969)$ & $1972$ & Co72 \\
       &              &             &            &        &    &    \\
B2.3   & $0.25$       & $3227$     & $3 \times 10$ & $+34^{\circ}
02^{\prime} < \delta < +40^{\circ}18^{\prime} (1969)$ & $1973$ & Co73 \\
       &              &            &             &         &   &   \\
B2.4   & $0.6$        & $448$   & $3 \times 10$ & $+21^{\circ} 40^{\prime}
 < \delta < +24^{\circ} 02^{\prime} (1969)$  & $1974$ & Fa74 \\
       &              &            &              &        &    &   \\
B3     & $0.1$        & $13354$  & $3 \times 5$ & $+37^{\circ} 15^{\prime} 
< \delta < +47^ {\circ} 37^{\prime} (1978)$ & $1985$ & Fi85 \\
       &              &           &             &    &   &   \cr
\hline
\end{tabular}
\end{center}
\end{table*}
The B3.1 survey has been completed with the same 
instrumental settings as for the previous B3 survey 
(Ficarra et al. 1985) but with a different observing and data reduction 
procedure. 
In the previous B2 Bologna surveys (Colla et al. 1970) 
the problems of reduced principal 
response and of grating response were solved by limiting the beam 
synthesis to the central part of the primary beam. This procedure 
however, requires a large number of discrete drift scans to cover 
a wide declination interval and is more demanding in observing time. 
In the B3 survey, the problems of reduced principal response and 
of grating responses were solved by recontructing the primary beam 
$BP(\delta)$ for any value of $\delta$, from a discrete sampling 
of the primary beam itself ($BP(\delta_{p})$), obtained by drift scans spaced 
$30$ arcmin. 
An interpolation with a SINC function made with 
$8$ coefficients revealed to be a good compromise 
between accuracy and computational effort. As a consequence, the first 
and the last four drift scans were lost in the interpolation process.
In case of the B3.1 survey, 
it would have been impossible to acquire four scans southward the 
survey limit $\delta = -2^\circ 00^\prime$, as this is the southernmost 
declination our telescope can observe. 
Thus, we went back to acquiring more closely spaced scans as in the B2 
surveys. 
At Dec. $\sim 0^{\circ}$ the 'Croce del Nord' transit 
radiotelescope has a $3^\prime \times 7^\prime$ HPBW and a sky strip 
$\sim 45^\prime$ wide in Declination is synthesized in a single 
scan. 
Drift scans were obtained spaced by $15^\prime$, allowing a considerable 
overlap between them. 
The observations were accomplished between March 1997 and June 1997. 
Many sky strips have been reobserved because of solar or man-made 
interference and poor weather conditions. 
Quality data checks were performed daily at the end 
of each observation by visual inspection of the calibration source 
parameters and the overall trend of 
the total power channels of the sky strip record. 
If considerable gain fluctuations or interferences were found, the 
observation was repeated. 

\noindent
The large overlap between contiguous scans yields at least two (or more, if 
the strip has been reobserved for any reason) independent measurements 
for any source, taken with different instrumental operating conditions. 

\noindent
This permits a weighted mean of the source parameters 
(sky coordinates, flux, fit residual) to be calculated. 
Furthermore, previous observations (Grueff et al. 1980) 
carried out with the same instrument 
of 358 sources in the $\pm 4^{\circ}$ strip of the Parkes $2700$ MHz 
survey allow important flux density checks to be done. 

\noindent
As a flux calibrator the source 3C 409 was used, with an adopted 
$S_{408} = 45$ Jy (see also Sect. 4) 
and it was observed before the beginning 
of each strip and after its end. The survey coordinates limits 
are: $-2^\circ 00^\prime$ to $2^\circ 15^\prime$ in Dec. and $21$h $00$m 
to $24$h $00$m, $00$h $00$m to $17$h $00$m in R.A. (J2000),
equivalent to $0.388$ sr. However, many sources have been extracted down 
to Dec. $= -2^\circ 15^\prime$ and up to Dec.$= +2^\circ 30^\prime$, but in 
these extended zones the catalogue is incomplete, as discussed later. 
All the results presented in this paper were obtained considering only the 
sources within the restricted declination range. 
The positions and flux densities were measured by a least-squares 
fitting of the beam function. 
The synthesis antenna pattern is described as follows: 

\begin{equation}
B(\alpha ,\delta) = BEW(\alpha, \delta) \cdot BNS(\delta)\, 
cos(K \alpha + \varphi)
\end{equation}

where $\alpha$ and 
$\delta$ are the usual astronomical coordinates; BEW and 
BNS represent the beam function in R.A. and Dec. respectively, and 
$\varphi$ represents the phase term between the E-W and the N-S arms 
due to their structure. 
The E-W and N-S arms beam functions are thus described:

\vspace{4mm}
\begin{equation}
BEW(\alpha, \delta) = 
{ {{sin(A\, \alpha)}\over{(A\, \alpha)}} \cdot 
{{sin \left( {{A} \over{24}} \alpha \right) } \over{{A}\over{24} 
\alpha}}}
\end{equation}


\vspace{4mm}
\begin{eqnarray}
BNS(\delta) = 
 {{sin[ A (cos (\delta_{0} - \gamma)) ( \delta - \delta_{0})]
} \over {8\, sin [ 
{{A} \over{8}}(cos (\delta_{0} - \gamma)) (\delta - \delta_{0})] }} 
\cdot  \nonumber\\
{{sin [ 
{{A}\over{8}}((cos(\delta_{0} - \gamma)) 
(\delta_{p} - \delta_{0}) ] } \over {
 8\, sin[ {{A}\over{64}}(cos(\delta_{0} - \gamma)) 
(\delta_{p} - \delta_{0})] }} 
\cdot \nonumber\\
e^-{ {(\delta_{p} - \delta_{0})^{2}}
\over {\Delta}}
\end{eqnarray}

\vspace{4mm}
In the above formulas, $\gamma = 44^\circ 34^\prime 21''$ is the Zenith of 
the North-South arm, $\delta_{0}$ is the source declination, 
$\delta_{p}$ is the ''primary beam'' pointing. Assuming $\alpha$ in units 
of $4$ seconds of time, and 
$(\delta - \delta_{0})$, $(\delta_{p} - \delta_{0}$) in units of $2$ 
arcminutes, the constants $A$,$K$,$\Delta$ are: $A = 1.5918$, $K = 0.74936$, 
$\Delta = 1152$. The phase term $\varphi$ is critical because it varies 
with the 
environmental and pointing changes, and it mostly contributes to the 
source position error in R.A. (an error of $1^{\circ}$ in this phase 
produces a shift of about $1.4''$ in R.A.). We decided to treat it 
as a free parameter to be fitted together with the source position and 
flux. 

The source extraction algorithm worked on each individual map synthetized in 
a drift scan. A beam function $(1)$ was fitted at each map maximum exceeding 
a $0.1$ Jy value. The fitting algorithm gave position, flux, phase term 
$\varphi$ 
and a fit r.m.s. residual. The fitted source was retained if the ratio 
flux/residual was larger than $4.6$. It is clear that, by accepting
sources 
with inferior ratio, we increase the chance of including spurious sources 
caused by noise or sidelobe confusion, while requiring a higher ratio 
could 
reject real sources. The ratio $4.6$ was chosen by comparing the resulting 
source counts (Log N - Log S) with the accurate determination of the 
Log N - Log S of the B3 survey (Grueff 1988). 

%
\begin{table*}
\begin{center}
\caption[]{The differential B3.1 Log N - Log S, compared 
to that of the B3 survey. For clarity, the convention adopted for 
the flux interval boundaries, is shown for the first interval only}
\begin{tabular}{cccccl}
\hline
Flux Interval & B3.1 counts & B3.1 - av.area & B3 counts & 
B3 - av.area & B3.1/B3 (-av.area) \\
\hline
              &     &     &      &       &      \\
$[0.1 - 0.15[$ & 1059 & 1001 & 2161 & 1816 & 0.55 \\
0.15 - 0.2   & 1119 & 998 & 1184 & 995  & 1.00 \\
0.2 - 0.3    & 1141 & 981 & 1198 & 1007 & 0.97 \\
0.3 - 0.5    & 894  & 788 & 981  & 825 & 0.96  \\
0.5 - 0.7    & 325  & 295 & 335  & 282 & 1.05 \\
0.7 - 1.0    & 210  & 194 & 236  & 198 & 0.98 \\
1.0 - 1.5    & 164  &     & 166  &     & 0.99 \\
1.5 - 3.0    & 92   &     & 114  &     & 0.81 \\
3.0 - 5.0    & 35   &     & 27   &     & 1.30 \\
5.0 - 10.0   & 12   &     & 13   &     & 0.92 \\
            &      &     &      &     &       \cr
\hline
\end{tabular}
\end{center}
\end{table*}

To obtain the Log N - Log S function, 
an 'avoidance area' has been defined as in Grueff (1988), to eliminate 
all areas possibly affected by sidelobes of strong sources. 
The sources with 
$S_{408}< 1$ Jy which fall inside a flux-dependent cross-shaped area, 
centered on each source with $S_{408} \geq 1$ Jy, have been flagged and 
excluded. 
An additional square area was excluded around those source with 
$S_{408} \geq 5$ Jy. The total avoided area was $0.062$ sr. 
Table 2 shows a comparison between the B3.1 and B3 
differential Log N - Log S. 
Column 1 contains the flux interval, columns 2 and 4 the sources counts 
(those of B3 scaled to the lower B3.1 area), 
columns 3 and 5 the sources counts after the flagged sources in the 
avoidance areas have been subtracted, and column 6 the ratio of the 
number counts in columns 3,5 (columns 2,4 for $S_{408} > 1$ Jy). 
The B3 survey 
is statistically complete to $S_{408} = 0.1$ Jy and sources down to 
$70$ mJy have been reliably counted outside the avoidance zones. To show
the amount of incompleteness in the B3.1 survey, the sources 
counts ratio in the flux interval $0.1 \leq S_{408} < 0.15$ Jy is reported 
in Table 2. The B3.1 survey shows an evident and considerable loss of 
sources at $S_{408} < 0.15$ Jy with respect to the B3 survey. 
The main cause of this incompleteness is the fact that a threshold of 
$0.1$ Jy for the pixel value was stated for the fitting algorithm. 
Table 3 shows the mean value of the $S/N$ 
of the sources (for $S_{408} < 1$ Jy, sources in the avoidance areas 
have been excluded) calculated for each flux interval. 
The $S/N$ values in Table 3 are mostly due to confusion, which
amounts to about $20$ mJy r.m.s. ( see Ficarra et al. 1985 for 
details ). The $S/N$ for strong sources is smaller than implied by this 
confusion error, because it includes an effect due to sources angular 
size. For instance, in the $3 - 5$ Jy interval, a $20$ mJy confusion error 
would imply a $S/N$ ratio of $200$; the observed lower value of $47.6$ 
is caused by the fact that several sources in this flux interval are 
sufficiently extended to give comparatively high residuals when fitted 
with a point-like model. Note that, as shown in Ficarra et al.(1985) 
(Table 3), confusion errors in the ''Croce del Nord'' give rise only to 
random flux errors and no systematic error. 
As the mean $\sigma$ for the lowest flux interval 
is $6.5$, we decided not to take into account any correction 
as in Bennet (1962), and to use the two surveys raw number counts (see 
also Murdoch et al. 1973). 
We conclude that the new B3.1 catalogue is statistically 
complete down to $0.15$ Jy with $3999$ sources; the number 
increases up to $5058$ at $S_{408} \geq 0.1$ Jy. 

\noindent
If the entire observed declination range is considered ($-2^\circ 15^\prime 
\leq \delta \leq +2^\circ 30^\prime$), the number of sources increases to 
$5578$ at $S_{408} = 0.1$ Jy, but the catalogue is no longer complete. 
In fact, the nominal lowest and highest declination pointings for the 
center of the strips are 
$\delta_{p} = -2^\circ 00^\prime$ and $+2^\circ 15^\prime$ respectively, 
and the reduced instrument response, due to the primary beam shape, causes the 
faintest sources to be lost near the sky strips declination boundaries. 
%
%
\begin{table*}
\begin{center}
\caption[]{The mean value of the $S/N$ for the B3.1 sources}
\begin{tabular}{cccc}
\hline
Flux Interval (Jy)& B3.1 - av.area & $< S/N >$ & $\sigma_{S/N}$ \\
\hline
              &      &     &      \\
$[0.1 - 0.15[$  & 1001 & 6.5 & $<0.1$ \\
0.15 - 0.2    & 998 & 7.7 & $<0.1$ \\
0.2 - 0.3     & 981 & 10.0 & 0.1 \\
0.3 - 0.5     & 788 & 13.8 & 0.2 \\
0.5 - 0.7     & 295 & 20.2 & 0.5 \\
0.7 - 1.0     & 194 & 25.6  & 0.9 \\
1.0 - 1.5     & 164 & 34.3  & 1.2 \\
1.5 - 3       & 92  & 39.2 & 2.0 \\
3 - 5         & 35  & 47.6  & 3.4 \\
5 - 10        & 12  & 54.5 & 8.6 \\
              &     &      &   \cr
\hline
\end{tabular}
\end{center}
\end{table*}
\smallskip
\section{The identification with the NVSS sources}
To obtain a good measurement of spectral index for the B3.1 sources, we 
used the NVSS sky survey data (Condon et al. 1998). These data are
available 
both as calibrated sky maps, with pixel size of $15''$ and resolution of 
$45''$ HPBW, and as a catalogue of sources. We preferred to use the map 
material, because it allows a better control of the problems arising in the 
cross-correlation of source catalogues produced at different frequencies 
and angular resolutions, and differently affected by incompleteness. 
For instance, if a $0.1$ Jy B3.1 source is not found in the NVSS
catalogue 
(which goes down to $2.5$ mJy), this implies one of the following: 
i$)$ a spectral index 
steeper than about $-3$; ii$)$ a spurious B3.1 ''source'', namely 
a sidelobe or the product of a blend of unrelated sources, due to the 
relatively poor angular resolution at $408$ MHz; iii$)$ incompleteness 
of the NVSS maps ( blanked areas or ''holes'' ). 
We found that, in the version at our disposal, about $7.3\%$ of map area 
was missing, i.e. the relative pixels were 'blanked'. Of course, this 
information is missing in the NVSS source catalogue, and to account for it 
correctly requires to use the map material. We thus developed our own 
source-measuring algorithm as follows. 

\noindent
A $15 \times 35$ pixels matrix is defined around the map pixel corresponding 
to the B3.1 source position. This rectangular area covers essentially 
all the B3.1 beam area. A two-component gaussian fit is performed, measuring 
the stronger source on the map, and approximating the two components 
with point-like sources. 
If this fitting gives a source separation less than $15''$, a single 
point-like component 
gaussian fit is preferred, and no radio size is given in this case. 
In each case the total source flux is computed and 
used to obtain the spectral index. However, if one or more NVSS pixels are 
'blanked' (henceforth BP) the whole procedure is skipped, the B3.1 source is 
marked as BP and it is excluded from further consideration. The statistics of 
the cross-ID is reported in Table 4.  
%
\begin{table*}
\begin{center}
\caption[]{Description of the NVSS cross-ID results}
\begin{tabular}{ll}
\hline
                              &            \\
B3.1 sources searched in NVSS & 5578 ($S/N \geq 4.6$) \\
B3.1 sources found            & 5269 \\
B3.1 sources with bad NVSS fit & 83 \\
$\theta \geq 45''$              & 1380 \\
$\theta < 45''$                 & 3299 \\
BP sources                      & 590  \\
B3.1 sources not found in NVSS & 226 \\
                &             \cr
\hline
\end{tabular}
\end{center}
\end{table*}
\section{Data quality checks}
In the following, checks about the quality of the $408$ MHz data will 
be discussed. It must be noticed however that, though the B3.1 sample has 
been defined by extracting the sources at the $4.6 \sigma$ level, we 
performed the checks using only sources with $S/N \geq 5.2$. 
This implied excluding $9 \%$ of the sources in the lowest complete flux 
interval considered ($0.15 - 0.3$Jy), but permitted to eliminate most of 
the spurious $408$ MHz sources. In the higher flux intervals, the fraction 
of the excluded sources is negligible. 

\subsection{Position uncertainties}
The B3.1 radio positions were used to perform the cross-ID with the NVSS 
maps. Because of the much larger HPBW 
of our instrument with respect to those of the VLA-NVSS, the 
$408$ MHz positions are certainly affected by much larger errors. 
The NVSS nominal 
r.m.s. position error is $\sigma_{P} \leq 1''$ for strong sources 
($S_{1400} \geq 15m$ Jy). As we considered sources with $S_{408} \geq 
0.15$ Jy even the sources with the steepest spectra have $S_{1400} \geq
20$ mJy. 
Thus the B3.1 position errors can be estimated by direct comparison with 
the NVSS positions. 
Only the NVSS unresolved sources ($\theta < 45''$) were considered, to 
avoid possible biases due to the source extension. 
Figs. 1 and 2 show the histograms of the position differences 
in the sense B3.1 - NVSS in rigth ascension and declination. 

\noindent
The distributions are reasonably represented by gaussian functions. 
Note however that expecially in the lower flux bins ($S_{408} < 0.5$ Jy), 
the distributions may be considered as the result of two different 
gaussian 
components: a relatively narrow gaussian plus a larger gaussian,
constituted of those sources with a poor $408$ MHz fit and consequent 
less precise low-frequency positions. 
For strong sources ($S_{408} > 0.5$ Jy) this effect is 
smaller as the increasing $S/N$ permits a better position measurement. 
The dispersion in the position differences is 
entirely attributable to the B3.1 measurements, and it was estimated 
as the $68$ percentile of the distribution. 
The systematic offsets contained in our positions were calculated 
as the median values of the coordinates displacements. 
Numerical data from Figs. 1 and 2 are listed in Table 5. 
Columns 3,5 contain the dispersion in R.A. and declination. 
The R.A. and Dec. offsets are reported in columns 4,6. All the 
values are in arcsec. 
%
\begin{table*}
\begin{center}
\caption[]{Differences between radio positions (B3.1 - NVSS). 
Positional errors are in arcsecs}
\begin{tabular}{cccccc}
\hline
Flux interval (Jy)& \# sources & R.A. $\sigma$ & R.A. offset & 
Dec. $\sigma$ & Dec. offset \\
\hline
              &      &        &                 &       &    \\
$[0.15 - 0.3[$  & 1270 & $17.8$ & $1.5_{-0.5}^{+0.4}$ & $19.3 $ 
& $2.3_{-0.4}^{+0.7}$ \\
0.3 - 0.5 & 625 & $12.7$ & $0.3_{-0.6}^{+0.5}$ & $13.4$ 
& $3.3_{-0.5}^{+0.4}$ \\
0.5 - 1.0 & 411 & $8.7$ & $-0.8_{-0.2}^{+0.5}$ & $9.1$
& $2.2_{-0.3}^{+0.4}$ \\
1.0 - 10.0& 245 & $5.9$ & $-0.6_{-0.4}^{+0.3}$ & $6.5$
& $2.6_{-0.3}^{+0.6}$ \\
              &      &        &                 &        &  \cr
\hline
\end{tabular}
\end{center}
\end{table*}
The cause of the small but significant declination offset ($\sim 2.6''$) 
is unknown, but is probably due to the large declination difference between the
sky survey region and the calibrator source used. 
The reason why the R.A. errors are similar to the Dec. errors lies in
the fact that, differently to the N-S direction, the interferometer E-W
instrumental response contains a term which is subjected to daily 
phase drifts, as shown in Ficarra et al.(1985). This results in an 
augmented positional error in the E-W direction, which almost equals 
that in the N-S direction.

\subsection{The B3.1 flux scale}
As mentioned in Sect. 2, the B3.1 primary flux calibrator was 3C 409, 
with an assumed $S_{408}= 45$ Jy. We performed some checks to establish 
the relationship between our and other flux scales from literature. 
First, a comparison with fluxes from previous observations 
(Grueff et al. 1980; hereafter G80) was made. In G80, the 
'Croce del Nord' observations of a sample of $358$ sources in the 
$\pm 4^\circ$ declination strip of the Parkes $2700$ MHz survey are 
described. 
The source 3C 123 was used as flux calibrator with an assumed 
$S_{408}= 120$ Jy ($2\%$ higher than Baars et al. 1977 scale; 
hereafter Ba77). 
We considered all the $26$ sources with $S_{408} \geq 3$ Jy in G80. The 
mean ratio B3.1/G80 was $0.98 \pm 0.01$. The inclusion of 
sources with $S_{408} \geq 1$ Jy in G80, gave a sample of $78$ sources, 
and a mean ratio B3.1/G80 of $0.99 \pm 0.01$ (see Fig.3). 
Instrumental non-linearity 
factors (receivers, correlators) were estimated not to exceed $1\%$ at the 
flux level of 3C 123. 
Thus we can state that the B3.1 flux scale is in accordance with Ba77 
to within $1\%$. 

\noindent
The $408$ MHz flux we adopted for 
3C 409 flux was derived from past 'Croce del Nord' observations of a set
of 3C sources taken from Riley (1988), whose fluxes were scaled to that of 
3C 123. 

\noindent
Another flux scale consistency check was made by comparing the B3.1 
sources fluxes with those taken from the $408$ MHz MRC sample 
( Large et al. 1981 ). By positional cross-coincidence, $500$ B3.1 
sources were found to be common to the MRC sample down to its limiting 
flux of $S_{408} = 0.7$ Jy. Only the MRC unresolved sources were 
retained. The mean ratio B3.1/MRC was 
$0.894 \pm 0.005$ ( see Fig.4 ). For the $45$ sources with $S_{B3.1} > 3$ Jy 
, a slightly better accordance was found, with a mean ratio 
B3.1/MRC $= 0.921 \pm 0.012$. If the sources with $S_{B3.1} \leq 3$ Jy 
are considered, the mean ratio B3.1/MRC is $0.891 \pm 0.005$. A nominal 
difference of $3\%$ is expected to be between the Ba77 flux scale 
and the Wyllie (1969) flux scale to which the MRC sample refers 
( Wy69/Ba77 $= 1.03$ ). Our data show a flux-scale discrepancy of about 
$11\%$, the Molonglo fluxes being too high of about $8\%$. 
This difference is similar, though less evident, to that found by 
Fanti et al. (1981) by comparing the B2 flux-scale ( $2\%$ below
Ba77 ) with that of the MC2 and MC3 ( Sutton et al. 1974 ) catalogues. 
The mean ratio B2/MC2 found was $0.851 \pm 0.006$ and was mostly 
attributed to an error in the Molonglo flux-scale. The same explanation 
was given by Grueff (1988) to justify why the Log N - Log S counts from 
MC2 and MC3 were found to be substantially higher than those from the 
B2 and B3 surveys. 

\noindent
We also considered the possibility of a daily instrumental gain drift 
to be present in our data, since each observing run lasted $20$ hours. 
For such a test to be performed, literature $408$ MHz fluxes of many 
sources uniformly distributed over the R.A. range covered, would be 
necessary. As low-frequency data are only available for a limited number 
of strong sources, we used two alternative methods. The first 
one consists in subdividing the whole R.A. range into a number 
of bins to compare the relative sources number counts. We considered 
sources with $S_{408} \geq 0.15$ Jy, $S/N \geq 5.2$ and R.A. boxes $5$
hours wide ( see Table 6 ). The obvious assumption of an isotropic sources 
distribution on the sky is made. 
We obtained an average of $941$ sources per box with a r.m.s. variation 
of $22$ sources, similar to the expected value of $31$ ($\sqrt{941}$), 
indicating a good isotropy and no 
systematic error affecting flux measurements. 

\noindent
Given a $B3$ integral Log N - Log S slope of $1.06$ in the density flux 
range $0.125 - 0.4$ Jy ( the slope for the B3.1 is $1.01$ in the 
$0.175 - 0.4$ Jy flux density interval ), a $5\%$ error in flux would imply a
variation in the number of sources of about $34$ units. 

The other method to check for instrumental gain fluctuations 
is based on the assumption that they give rise to spectral index 
fluctuations of the sources (see next section). 

\subsection{Spectral indices}
The spectral index distribution of the B3.1 sources is shown in Fig.5. 
We can use the spectral index distributions obtained in different R.A. 
intervals to check for instrumental gain fluctuations. 
The survey right ascension interval was subdivided into $4$ boxes, five 
hours wide, and the median spectral index was calculated for each box. 
We considered those sources with $S_{408} \geq 0.15$ Jy, $S/N \geq 5.2$ 
which have an unresolved counterpart in the NVSS. 
Spectral data are also reported in Table 6. 
The median values of $\alpha_{408}^{1400}$ differ up to the $2\sigma$ 
level, indicating a marginal variation with the right ascension. This is 
substantially in agreement with the result of Sect. 4.2, showing no 
noticeable gain change with the right ascension. 
A change of $\pm 1\%$ in instrumental gain would produce a change of 
$\mp 0.008$ in spectral index.

%
\begin{table*}
\begin{center}
\caption[]{Median spectral indices for the $4$ R.A. intervals}
\begin{tabular}{ccc}
\hline
R.A. & \# sources & median $\alpha_{408}^{1400}$ \\
\hline
            &     &                       \\
$[21h - 02h[$ & 928 & $-0.80_{-0.005}^{+0.015}$ \\
02h - 07h   & 934 & $-0.77_{-0.010}^{+0.005}$ \\
07h - 12h   & 946 & $-0.76_{-0.010}^{+0.005}$ \\
12h - 17h   & 956 & $-0.81_{-0.005}^{+0.010}$ \\
            &     &                           \cr
\hline
\end{tabular}
\end{center}
\end{table*}

\noindent
The existence of a correlation between the flux density and the spectral 
index could be relevant to the study of the radio sources evolution. 
Previous works (Grueff et al. 1995 and references therein) did not show 
any variations of the average spectral index with the flux 
density in the range $0.03 < S_{408} < 4Jy$. We subdivided the flux range 
$0.15 < S_{408} < 10Jy$ into four bins and we calculated 
the median spectral index for each flux bin. 
The straight line (see Fig.6) is the weighted least-square fit to 
the data points and the following relation resulted: 

\[ \alpha_{408}^{1400} = (-0.7697 \pm 0.0060) - (0.0197 \pm 0.0120)~
\log~S_{408} \]

\noindent
with a $\chi_{\nu} = 0.68$ . However a fit with an horizontal line gives 
$\chi^{2} \sim 4.5$, corresponding to $20\%$ probability. 
Thus, we find little evidence of a change in spectral index with the flux. 
\section{The B3.1 USS sample}

\subsection{The sample definition}
The selection criteria defining our USS sample are: $S_{408} \geq 0.1$ Jy, 
$S/N \geq 6.5$, $\theta < 45''$ ( as measured on the NVSS maps ) 
and $\alpha_{408}^{1400} < -1$. 
Furthermore, the zone with R.A $05$h $30$m to $08$h $30$m was excluded 
to avoid low galactic latitude objects (i.e. $|b| < 15^\circ$). 
Out of a list of $5578$ sources, only about $3.3\%$ were retained 
after applying the selection criteria above, and accurately checking 
the individual radio data. The B3.1 USS sample thus contains $185$ sources. 
The numerical values adopted for the 
selection criteria are similar to those used by other groups 
(Blundell et al. 1998; de Breuck et al. 1997; Rhee et al. 1996). 
The angular size criterium helps to eliminate relatively nearby objects, 
objects with $1.4$ GHz flux affected by resolution errors, and extended 
sources dificult to identify optically. 
There are $2601$ B3.1 sources with a NVSS cross-ID, $S/N \geq 6.5$ and 
$\theta < 45''$ (hereafter B3ID sample) and these provided 
the $185$ USS sample. 
In the following we try to estimate the number of USS lost because of 
the various sources of incompleteness. 
There are two important factors which may affect the B3.1 USS 
sample. 

First, the B3.1 survey is statistically complete 
to $S_{408} = 0.15$ Jy, but the USS have been extracted down to 
$S_{408} = 0.1$ Jy. From Table 2 (columns 2,4 since for the USS search 
the avoidance areas have not been considered) and Fig.7 we 
calculate that about $13$ USS have not been detected in the flux bin 
$0.1 - 0.15$ Jy because of the B3.1 incompleteness. 

Secondly, the presence of BPs on the NVSS maps does not 
permit the spectral index for a considerable fraction of B3.1 sources 
to be calculated. Contrarily to the previous one, this 
factor affects all the USS population, irrespectively of the radio 
flux. There are $438$ BP sources with $S/N \geq 6.5$. As 
$71 \%$ of the sources with a NVSS cross-ID and $S/N \geq 6.5$ 
have $\theta < 45''$, then out of $438$ BP sources, 
$311$ ($438 \times 0.71$) are expected to have $\theta < 45''$. 
We expect that out of these $311$ BP sources, $25$ 
could be USS lost because of BPs in the NVSS maps. 

To conclude, we estimate that about $38$ ($13+25$) USS have not been 
detected because of these incompleteness factors. 

\subsection{The B3.1 USS radio properties}

As a first approximation, probing lower fluxes should correspond to sampling 
more distant objects. Although this zero-order expectation 
could turn out to be wrong, 
our survey has been made following this suggestion. 
To compare the relative sentitivities of the most important low-frequency 
radio surveys, used to derive USS samples, a plot of the limiting flux 
versus selection frequency is shown in Fig.7. 
With respect to the 4C USS sample (Chambers et al. 1996), 
the B3.1 USS allows to investigate 
sources almost one order of magnitude deeper in flux, resulting in a lower 
intrinsic radio power or possibly a higher redshift 
for the objects. Moreover, as a radio-optical correlation 
exists (Tielens et al. 1979, Laing \& Peacock 1980), probing low radio 
powers helps to select those objects with low optical luminosity as well. 
Their optical emission is thus less contaminated by the 
AGN nucleus and the surrounding stellar population can be better 
outlined. Furthermore, a wider coverage of the $P-z$ plane is achieved,
with benefits in the statistical study of the radio sources intrinsic 
properties. Fig.8 shows the $S_{408}$ distribution for the 4C USS 
and our sample. 
For the 4C USS, the $S_{408}$ was obtained by the 
$\alpha_{178}^{1415}$ or taken from the B3 catalogue when available. 
The lack of USS with $S_{408} < 1$ Jy is clearly seen in the 
4C sample. 
Note that only $9 \%$ of the B3.1 USS have $S_{408} \geq 1$ Jy. 
The most recent USS sample, selected by a low-frequency survey and also 
covering the same sky region as our, is the TN sample (de Breuck et al. 
1997). It resulted from a cross-correlation of the TEXAS catalogue 
with the NVSS, adopting a spectral index cut-off $\alpha_{365}^{1400} 
< -1.3$. As the TEXAS survey ( Douglas et al. 1996 ) 
is $\sim 80\%$ complete at 
$S_{365} = 250$ mJy, the B3.1 USS sample contains sources up to two
times fainter, and from Fig.8 we estimate that $\sim 50$ B3.1 USS have
fluxes too faint to have been included in the TN sample. 

\noindent
Fig.9 shows the spectral index distribution for the 4C USS and B3.1 
USS samples. 
The median $\alpha_{408}^{1400}$ is $-1.07 \pm 0.01$ while the median 
$\alpha_{178}^{1415}$ is $-1.12_{-0.03}^{+0.02}$. The two distributions 
are very similar. 

Although higher resolution maps are needed to measure precise radio sizes 
and positions, 
some statistical considerations may be derived also using the radio 
sizes given by our fitting algorithm. A comparison between 
the NVSS catalogue deconvolved sizes and the B3.1 radio sizes for a
subset of USS sources, showed a substantial agreement within few arcseconds. 
Out of the $185$ USS, $85$ were fitted with a 
double-component model, while the remaining $100$ sources (54\%) 
had overall sizes 
less than $15''$ and were fitted with a single-component model (see 
Sect. 3). 
Thus there is indication that about $50\%$ of the B3.1 USS are 
smaller than $15$ arcsec (see Fig.10). 
Other authors (Blundell et al. 1998) 
used much restrictive size selection criteria ($\theta < 15''$) 
with the aim of retaining the ${\it z} > 4$ USS objects in the 6C survey. 
However, they showed that about $44 \%$ of the 3C radiogalaxies, 
if redshifted to ${\it z} = 3$, would be lost because of this size cut. 
As for the known USS some spread in $\theta$ is evident, a too low size 
cut could make easier the identification process, but it may 
result in excluding several sources. 
We compared the size distribution of our USS with that of 
the existing $365$ MHz samples (R\"{o}ttgering et al. 1994), finding 
a good agreement. 
\subsection{The optical ID programme}
Because we are interested in selecting only very distant (and thus optically 
very faint) objects, we firstly examinated the 
digitized POSS-I red plates to select those USS with no visible optical 
counterpart. A square box $20$ 
arcsec wide centered at the NVSS position was inspected, resulting in a 
total of $146$ (79\%) empty fields (EF) down to $R \sim 20.0$. 
Any USS with at least one optical object falling inside the box was 
excluded, irrespectively of it's true optical ID. 
Consequently, $79\%$ is a lower limit to the fraction of EF in the USS
sample. 
The adopted box size value was derived from very conservative 
considerations about the NVSS r.m.s. position error, the possible 
intrinsic radio-optical displacement of a radiosource, and the NVSS 
maps low resolution. 

\noindent
High resolution ($1 - 2''$) maps and much deeper optical material 
are necessary for a secure optical 
identification. 
To date, the VLA-FIRST $1.4$ GHz (Becker et al. 1995) survey overlaps 
only in part 
the B3.1 survey region. We obtained FIRST maps for $50$ B3.1 
USS, and in most cases, the resolution ($5.4'' \times 6.4''$) is high 
enough to permit a morphological classification of the sources. 
For $8$ USS, the optical counterpart is visible on the POSS-I plates. 
For the remaining $42$ USS, the optical identification 
on the deeper POSS-II ($R = 20.8$) was carried out, resulting in  one ID, 
two uncertain cases and 39 are EF.
Out of these $42$ USS, $12$ are unresolved, $27$ 
are double, $2$ are triple and one is a multiple source. 
\section{Conclusions}
This paper describes the first results of an effort to discover 
high-z radiogalaxies selecting very steep spectrum radiosources.

The technique of the radio spectrum steepness has been exploited by many 
groups and it revelead to be an effective way to find high redshift 
radiogalaxies. 
The present USS sample is about one 
order of magnitude deeper in flux with respect to the 4C USS sample,
and about a factor of two with respect to the recent TN sample (van 
Breugel et al. 1998) based on the TEXAS $365$MHz survey. 

\noindent
In this paper we described the realization of the survey, which was the 
first step to build-up the B3.1 USS sample, and some details of the 
observing and reduction procedure. 
The definition of the B3.1 sources catalogue was 
obtained by comparison of the Log N - Log S function with that of the 
previous B3 survey. The B3.1 survey is complete to $0.15$ Jy but many 
sources down to $0.1$ Jy are listed giving a total of $5058$ sources. 
A cross-ID with the NVSS $1.4$ GHz maps gave us precise radio positions
and fluxes to calculate the spectral index $\alpha_{408}^{1400}$. 
The radio positions displacement (B3.1 - NVSS) histograms have 
gaussian shape with a dispersion decreasing from about $15''$ in the flux 
interval $0.15 - 0.3$ Jy to about $5''$ in the $1 - 10$ Jy interval. 
Our flux scale is in accordance with that of 
Baars et al. (1977) to within $1\%$, while daily instrumental 
gain drift 
possibly affecting our data were found to be negligible. 
We found no evidence of a change of the radio spectrum as flux decreases. 
As USS selection criteria we used: 
$\alpha_{408}^{1400} < -1$ and $\theta < 45''$ and these gave 
a total of $185$ sources which constitute the B3.1 USS sample. 
We started the search of their 
optical counterparts on the digitized POSS-I plates in order to select 
only the empty fields to $R = 20.0$. 
Furthermore, we collected FIRST maps ($5.4'' \times 6.4''$ resolution) 
for $50$ USS sources and we searched for their optical counterparts on 
the POSS-II prints ($R = 20.8$). For $39$ USS no optical ID was found, and 
out of these, $12$ are unresolved in the FIRST maps. The fact that these 
small sources also have the steepest spectra among the B3.1 USS sources 
suggests they could reliably be high-z objects. 

\noindent
To date, VLA $4.85$ GHz observations have been carried out for 
a subset of $128$ B3.1 USS sources, giving $5 \times 4''$ HPBW radio 
maps and precise flux measurements. 
For $6$ of them, $K'$ band images have been acquired with the 
new italian telescope TNG during its testing phase. 
These results will be presented in a future paper. A preliminary version 
of the B3.1 Catalogue is available in digital form via anonymous FTP 
at the following address: ftp terra.ira.bo.cnr.it; 
cd /astro/B3.1
%
%

%
%
\clearpage
\begin{figure*}[p]
\vspace{23cm}
\includegraphics{H1564R.F1a}
\includegraphics{H1564R.F1b}
\includegraphics{H1564R.F1c}
\includegraphics{H1564R.F1d}
\caption{Differences between right ascensions ($B3.1 - NVSS$) in 
different intervals of flux (at $408$ MHz)}
\end{figure*}
%
\clearpage
\begin{figure*}[p]
\vspace{23cm}
\includegraphics{H1564R.F2a}
\includegraphics{H1564R.F2b}
\includegraphics{H1564R.F2c}
\includegraphics{H1564R.F2d}
\caption{Differences between declinations (B3.1 - NVSS) in different 
intervals of flux (at $408$ MHz)}
\end{figure*}
%
\clearpage
\begin{figure*}[p]
\vspace{20cm}
\includegraphics{H1564R.F3}

\caption{Flux densities (B3.1 vs G80) for $78$ sources; those with  
$S_{408} < 3$ Jy in G80 are marked with crosses}
\end{figure*}
%
\clearpage
\begin{figure*}[p]
\vspace{20cm}
\includegraphics{H1564R.F4}

\caption[]{Flux densities (B3.1 vs MRC) for $500$ sources; those with
$S_{408} \leq 3$ Jy in B3.1 are marked with crosses}
\end{figure*}
%
%
\clearpage
\begin{figure*}[p]
\vspace{20cm}
\includegraphics{H1564R.F5}

\caption{Spectral indices distribution for the B3.1 sources with 
$S/N \geq 4.6$ and $\theta < 45''$}
\end{figure*}
%
\clearpage
\begin{figure*}[p]
\vspace{19cm}
\includegraphics{H1564R.F6}

\caption{Median spectral index as a function of flux density}
\end{figure*}
%
\clearpage
\begin{figure*}[p]
\vspace{2cm}
\resizebox{14cm}{!}{\includegraphics {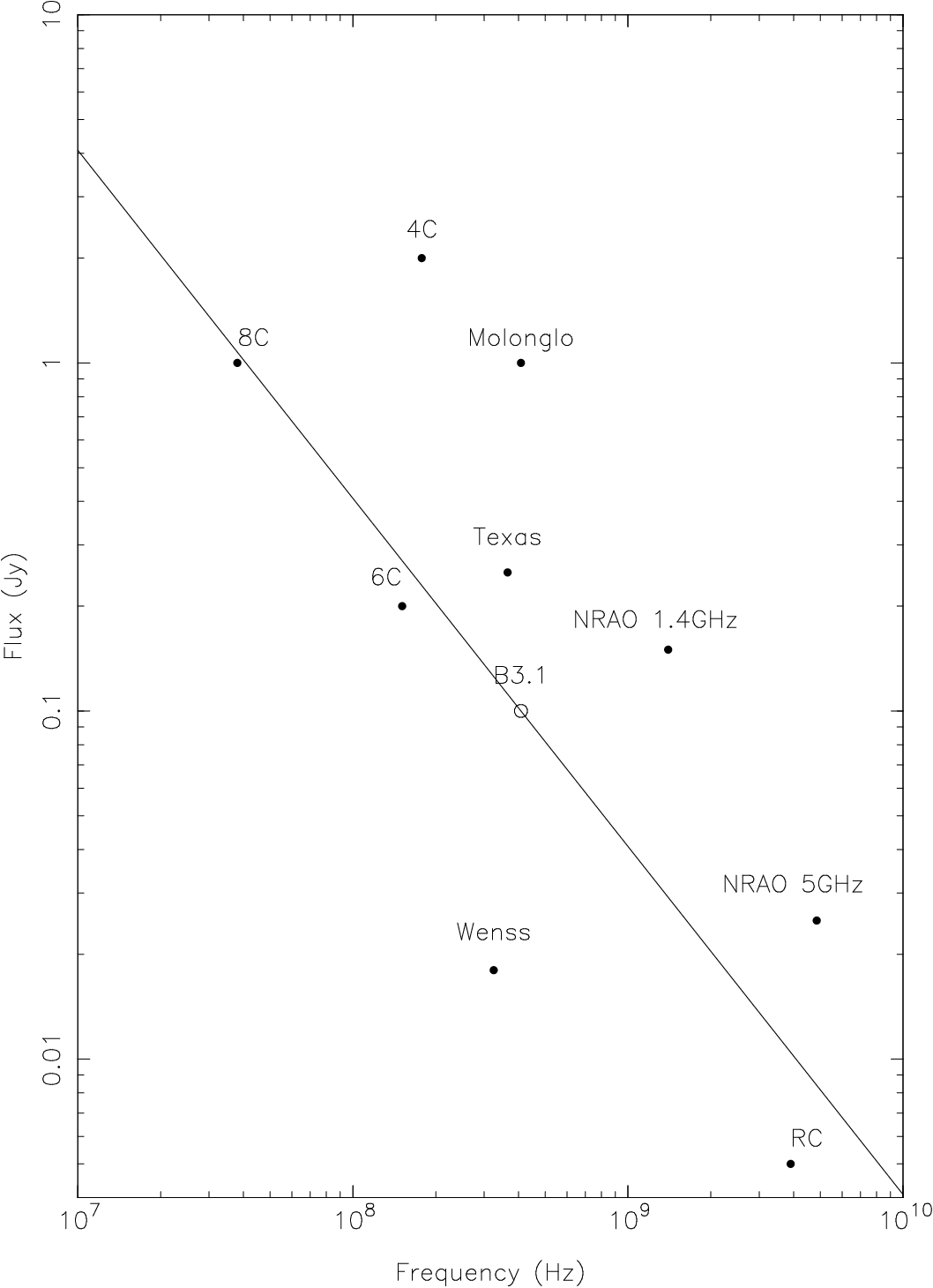}}

\caption{Low-frequency radio surveys used to derive USS samples. 
The straight line corresponds to a source with a spectral index 
$\alpha_{408}^{1400} = -1$, a value adopted in most of cases as a USS 
selection criterium}
\end{figure*}
%
\clearpage
\begin{figure*}[p]
\vspace{20cm}
\includegraphics{H1564R.F8}

\caption{The $S_{408}$ distribution for the B3.1 sample and 4C 
USS sources from Tielens et al. (1979) (only sources with 
$\theta < 45''$ were considered)}
\end{figure*}
%
\clearpage
\begin{figure*}[p]
\vspace{20cm}
\includegraphics{H1564R.F9}

\caption{The spectral index distributions for the B3.1 and 4C USS 
sample}
\end{figure*}
%
\clearpage
\begin{figure*}[p]
\vspace{20cm}
\includegraphics{H1564R.F0}

\caption{Distribution of the B3.1 USS sources radio sizes fitted on 
the NVSS maps. The dashed line encloses the sources with $\theta < 15''$ 
for which a single component fit was performed}
\end{figure*}
\end{document}